\documentclass[10pt,fleqn]{article}

\usepackage{amssymb}

\newcommand{\name}[1]{\begin{flushleft}
                       \LARGE \bf #1
                       \end{flushleft}\vspace{-3mm}}

\newcommand{\Author}[1]{\begin{flushleft}
                       \it #1 \end{flushleft}}

\newcommand{\Adress}[1]{\begin{flushleft}
                       \it #1 \end{flushleft}}

\newcommand{\AbsEng}[1]{
    \begin{flushright}
    \begin{minipage}{120mm}
     \small   #1
    \end{minipage}
    \end{flushright}
}

\newcommand{\be}{\begin{equation}}
\newcommand{\ee}{\end{equation}}
\newcommand{\ba}{\hspace*{-5pt}\begin{array}}
\newcommand{\ea}{\end{array}}
\newcommand{\p}{\partial}
\newcommand{\ds}{\displaystyle}

\begin{document}

\name{On representations of the inhomogeneous \\
de Sitter group and on
equations\\
 of the Schr\"odinger--Foldy type}

\medskip

\noindent{published as Preprint of Institute of Theor. Phys., N~69-1, Kyiv, 1969, 22~p.}

\Author{Wilhelm I. FUSHCHYCH and Ivan Yu. KRIVSKY}

\Adress{Institute of Mathematics of the National Academy of
Sciences of Ukraine,\\ 3 Tereshchenkivska  Street, 01601 Kyiv-4,
UKRAINE}

\noindent {\tt URL:
http://www.imath.kiev.ua/\~{}appmath/wif.html\\ E-mail:
symmetry@imath.kiev.ua}

\AbsEng{This paper is a continuation and elaboration of our works~[1] where some approach
to the variable-mass problem were proposed. Here we have found a concret realization of
irreducible representations of the inhomogeneous group $P(1,n)$ ---
the group of translations and rotations in $(1+n)$-dimensional Minkowski space in two
classes (when $P_0^2-P_k^2>0$ and $P_0^2-P_k^2<0$). All the $P(1,n)$-invariant equations
of the Schr\"odinger--Foldy type are written down. Some questions of a physical
interpretation of the quantum, mechanical scheme based on the inhomogeneous de
Sitter group $P(1,n)$ are discussed.

\hspace*{5mm}Report presented at the Conference on Composite
Models of Elementary Particles (Institute for
Theoretical Physics, Kiev, Ukrainian SSR, June 1968).}

\medskip

\centerline{\bf 1. Introduction}

Recall here the initial points of our approach to the variable mass problem proposed in ref.~[1]:

A. The square of variable mass operator is defined as an independent dynamical variables
\be
M^2\equiv \varkappa^2 +P^2_4,
\ee
where $\varkappa$ is a fixed parameter and $P_4$ is an operator lice the components of
three-momentum $\vec P$, which commutes with all the generators of the algebra
$P(1,3)$ of the Poincar\'e group.

B. The relation between the energy $P_0$,  three-momentum $\vec P$  and variable-mass
$M$ of a physical system is remained to be conventional (here everywhere $\hbar =c=1$):
\be
P_0^2 =\vec P^{\,2} +M^2\equiv P_k^2+\varkappa^2, \qquad k=1,2,3,4.
\ee

C. The spaced $p\equiv (p_0, p_1, \ldots, p_4)$ and $x\equiv (x_0, x_1, \ldots, x_4)$
are assumed to be plane and reciprocally conjugated. It follows then from A, B and C that
the generalized relativistic group symmetry is the inhomogeneous de Sitter
group\footnote{The algebras and groups connected with them are designated here with
the same symbols.} $P(1,4)$ --- the group of translations and rotations in five-dimensional
Minkowski space.  This group is a minimal extention of the conventional group
of relativistic symmetry --- the Poincar\'e group $P(1,3)$.

In this paper we shall present a further studying of the approach proposed in ref.~[1].
In particular, the main assertions which were briefly formulated in ref.~[1], are
generalized here and their detail proofs are given. In Section~2 a concrete realization of
irreducible representations for the generators $P_\mu$, $J_{\mu\nu}$ of the algebra
$P(1,n)$ with arbitrary $n$ carried out, which made it possible to give a proof of the
$P(1,n)$-invariance of the Schr\"odinger--Foldy type equations written flown in ref.~[1]
for $n=4$. Some questions of a physical interpretation of quantum mechanical scheme
based on the group $P(1,4)$ are considered in Section~3.

\medskip

\centerline{\bf 2. Realizations of the algebra representations}

For the sake of generality all the considerations are made here not for the de Sitter group
$P(1,4)$ but for the group $P(1,n)$ of translations and rotations in dimensional
Minkowski space, which leaves unchanged, the form
\be
\ba{l}
x^2\equiv x_0^2 -x_1^2-\cdots -x_n^2 \equiv x_0^2-x_k^2\equiv x_\mu^2,
\vspace{1mm}\\
\mu =0,1,2,\ldots, n; \qquad k=1,2,\ldots, n,
\ea
\ee
where $x_\mu$ are differences of point coordinates of this space.

Commutation relations for the generators $P_\mu$, $J_{\mu\nu}$ of the algebra
$P(1,n)$ are choosen in the form
\renewcommand{\theequation}{\arabic{equation}{\rm a}}
\setcounter{equation}{3}
\be
\left[ P_\mu, P_\nu\right]=0, \qquad
-i\left[ P_\mu, J_{\nu\sigma}\right]=g_{\mu\nu} P_\sigma-g_{\mu\sigma}P_\nu,
\ee
\renewcommand{\theequation}{\arabic{equation}{\rm b}}
\setcounter{equation}{3}
\be
-i \left[ J_{\mu\nu}, J_{\rho\sigma}\right] =g_{\mu\sigma} J_{\nu\rho} +g_{\nu\rho}J_{\mu\sigma}
-g_{\mu\rho}J_{\nu\sigma} -g_{\nu\sigma} J_{\mu\rho},
\ee
where $g_{00}=1$, $-g_{kl}=\delta_{kl}$, $P_\mu$ is Kroneker symbol, $P_\mu$
are operators of infinitesimal displacements and $J_{\mu\nu}$  are operators
infinitesimal rotations in planes $(\mu\nu)$.

Authors of refs.~[2--5] have studed all the irreducible representations of the Poincar\'e
group $P(1,3)$ and have found the concrete realization for the generators of its algebra.
Their methods we generalize here for the case of group $P(1,n)$. But all the treatments are
carried out in more general and compact form then it was done even for the case of $P(1,3)$.

For representations of the class I ($P^2\equiv P_0^2 -P_k^2>0$)
when the group $O(n)$ of rotations in a $n$-dimensional Euclidean space is the little group
of the group $P(1,n)$, the generators $P_\mu$, $J_{\mu\nu}$ are of the form
\renewcommand{\theequation}{\arabic{equation}}
\setcounter{equation}{4}
\be
\ba{l}
P=p\equiv (p_0, p_1,\ldots, p_n)\equiv (p_0, p_k),
\vspace{3mm}\\
J_{kl}=x_{[k}p_{l]}+S_{kl},
\qquad \ds J_{0k}=x_{[0}p_{k]}-\frac{S_{kl} p_l}{\sqrt{p^2+p_k^2}+\sqrt{p^2}},
\ea
\ee
where
\[
P^2\equiv p^2\equiv p_0^2-p_k^2>0, \qquad x_{[\mu}p_{\nu]} \equiv x_\mu p_\nu -x_\nu p_\mu,
\]
operators $x_\mu$, $p_\mu$  are defined by relations
\be
[x_\mu, p_\mu]=-ig_{\mu\nu}, \qquad [x_\mu, x_\nu]=[p_\mu,p_\nu]=0,
\ee
and $S_{kl}$ are matrices realizing irreducible representations $D(s,t,\ldots)$ of the algebra
$O(n)$ which have been completely studied in ref.~[6] (here the numbers $s,t,\ldots$
are numbers which identify a correspondence irroducible representations of the
algebra  $O(n)$).  Using~(6) and relations for the generators $S_{kl}$
(which are not written down here), one can immediately verify that~(5)
actually satisfy the relations~(4). Since in this class the little group of the group $P(1,n)$
coincides with the compact group $O(n)$, all the irreducible representations of the group
$P(1,n)$ are here unitary and finite-dimensional (concerning a set of ``spin'' indexes
$s_3,t_3, \ldots$).

A concrete form of the operators  $P_\mu$, $J_{\mu\nu}$ which are defined by Eqs.~(5),
depends on a choice of concrete form of matrices $S_{kl}$ and operators $x_\mu$,
$p_\mu$ which are defined by relations~(6). The concrete form of the operators $x_\mu$, $p_\mu$
and $S_{kl}$ depends on what of operators, constituting a complete set of commuting
dynamical variables are operators of multiplicationd (``diagonal operators''). The sets
$(P_0, P_1,\ldots, P_n, S_3, T_3, \ldots)$ or \linebreak $(x_0, x_1, \ldots, x_n, S_3, T_3, \ldots)$
are examples of such a complete sets where $S_3, T_3, \ldots$ are all the independent
commuting generators of the algebra $O(n)$. In the general case a complete set of dynamical
variables is constructed from the corresponding number of commuting combinations
of operators $x_\mu$, $p_\mu$ and $S_{kl}$. Different concrete forms of operators
$P_\mu$, $J_{\mu\nu}$ which are defined by the choice of other complete set as diagonal,
are connected by unitary transformations. The form~(5) for the generators is the most
general in the sense that it is not bound to the choice of concrete complete set as diagonal.

A few words about a apace of vectors $\Psi$, in which the operators~(5)
are defined. It is an Hilbert space of vector-functions depending on the eigenvalues
of operators of a diagonal complete sel. For instance, in the $x$-representation where
the operators $x_\mu$ are diagonal (i.e., are operators of multiplication) and,
as it follows from relations~(6), $p_\mu=ig_{\mu\nu} \p_\nu$, $\p_\nu\equiv \p/\p x_\nu$
the operators~(5) are defined in the Hilbert apace of the vector-function $\Psi=
\Psi(x)=\Psi(x_0, x_1,\ldots, x_n)$ of $(1+n)$ independent variables $x_\mu$.
The components of a vector $\Psi$  are functions not only of $x_0,x_1,\ldots,x_n$
but also of auxiliary variables $s_3,t_3,\ldots$, i.e., are functions
$\Psi(x_0,x_1,\ldots, x_n, s_3,t_3,\ldots)$,  where $s_3,t_3,\ldots$
are eigenvalues of ``spin'' operators $S_3,T_3,\ldots$ and, as it is known, take discrete values.
In $p$-representation where $p_\mu$ are operators of multiplication and, according
to~(6), $x_\mu=ig_{\mu\nu} \p/\p p_\nu$  vector-functions are $\Psi=\widetilde \Psi(p)
\equiv \widetilde \Psi(p_0, p_1,\ldots, p_n)$ and their components are
$\widetilde \Psi(p_0,p_1,\ldots, p_n,s_3,t_3,\ldots)$. The scalar product of vectors $\Psi$
is defined~as
\be
\ba{l}
\ds (\Psi, \Psi') \equiv \int d^{1+n} x\; \Psi^+(x) \Psi'(x)=
\vspace{2mm}\\
\ds \qquad =\int d^{1+n} x \sum_{s_3,t_3,\ldots} \Psi^* (x, s_3, t_3, \ldots) \Psi'(x,s_3, t_3,\ldots)=
\vspace{2mm}\\
\ds \qquad = \int d^{1+n} p\; \widetilde \Psi^+(p) \widetilde \Psi'(p)=
\int d^{1+n} p \sum_{s_3, t_3, \ldots} \widetilde \Psi^* (p, s_3,t_3,\ldots)
\widetilde \Psi'(p, s_3, t_3, \ldots),
\ea\hspace{-10.82pt}
\ee
where $d^{1+n} x=dx_0 d x_1 \ldots d x_n$, $\Psi$ and $\widetilde \Psi$ being connected
by Fourier-transformations.

For representations of the class III ($P^2 =P_0^2-P_k^2<0$) when the little group of the group
$P(1,n)$ is already uncompact group $O(1,n-1)$ of rotations in
$1+(n-1)$-dimensional pseudo-Euklidean space, the generators $P_\mu$, $J_{\mu\nu}$
are of the form
\be
\ba{l}
P=p\equiv (p_0,p_k)=(p_0, p_a, p_n),
\vspace{2mm}\\
\ds J_{ab}=x_{[a} p_{b]} +S_{ab}, \qquad
J_{an} = x_{[a} p_{n]} -\frac{S_{ab} p_b -S_{a0} p_0}{\sqrt{-p^2-p_a^2 +p_0^2} +\sqrt{-p^2}},
\vspace{2mm}\\
\ds J_{0a}=x_{[0} p_{a]} +S_{0a}, \qquad
J_{0n} = x_{[0} p_{n]} -\frac{S_{0a} p_a}{\sqrt{-p^2-p_a^2 +p_0^2} +\sqrt{-p^2}},
\ea
\ee
where $a,b=1,2,\ldots,n-1$ the operators $x_\mu$, $p_\mu$ are defined by
relations~(6) as before and the operators $(S_{0a}, S_{ab})$ are generators of the algebra
$O(1,n-1)$ in corresponding irreducible representations which have been well studied
by Gelfand and Grayev~[7].

Components of vector-functions, in the space of which the operators~(8)
are defined, are the functions of variables $s_3,t_3,\ldots$
(besides of variables $x_\mu$ or $p_\mu$, of course) which are the eigenvalues of the
corresponding independent commute generators of the algebra $O(1,n-1)$.
In contrast to the case~I, in this case the variables $s_3,t_3,\ldots$ may take both discrete
and continual valuea. Remind (see ref.~[7]) that the group $O(1,n-1)$
has both unitary and nonunitary representations, all the unitary representations being
infinite-dimensional (in the last case the ``spin'' variables $s_3,t_3,\ldots$
take continual values). In accordance with this, among the representations of the group
$P(1,n)$ in the class~III there will be both unitary (only infinite-dimensional) and nonunitary
(finite- and infinite-dimensional) irreducible representations.

Now we shall give here a recipe of constructing the representations of the class~III from
those of the class~I.

Note first that if operators $P$, $J$ realize representation of the algebra $P(1,n)$,
then operators  $\widetilde P$, $\widetilde J$ being defined by means of
\renewcommand{\theequation}{\arabic{equation}{\rm a}}
\setcounter{equation}{8}
\be
(P_0, P_a, P_n)=(-i\widetilde P_n, \widetilde P_a, i\widetilde P_0),
\ee
\renewcommand{\theequation}{\arabic{equation}{\rm b}}
\setcounter{equation}{8}
\be
\left(
\mbox{\begin{tabular}{c|c}
$J_{0a}$ & $J_{0n}$ \\[2mm]
\hline
&\\[-2mm]
$J_{ab}$ & $J_{an}$
\end{tabular}}\right)=
\left(
\mbox{\begin{tabular}{c|c}
$-i\widetilde J_{na}$ & $\widetilde J_{n0}$ \\[2mm]
\hline
&\\[-2mm]
$\widetilde J_{ab}$ & $i\widetilde J_{a0}$
\end{tabular}}\right),
\ee
realize a representation of the algebra $P(1,n)$ too. To proof this assertion, it is enough to
verify that from the commutation relations~(4) for $P$, $J$ and from definitions~(9),
it follows that the operators $\widetilde P$, $\widetilde J$satisfy the commutation relations~(4)
too.

Define, further, the operators $\widetilde x$, $\widetilde p$ and $\widetilde s$
by means of the following relations
\renewcommand{\theequation}{\arabic{equation}{\rm a}}
\setcounter{equation}{9}
\be
(x_0, x_a, x_n) =(-i\widetilde x_n, \widetilde x_a, i\widetilde x_0),
\ee
from
\renewcommand{\theequation}{\arabic{equation}{\rm b}}
\setcounter{equation}{9}
\be
(p_0, p_a, p_n) =(-i\widetilde p_n, \widetilde p_a, i\widetilde p_0),
\ee
\renewcommand{\theequation}{\arabic{equation}{\rm c}}
\setcounter{equation}{9}
\be
(S_{ab}, S_{an}) =(\widetilde S_{ab}, i\widetilde S_{a0}).
\ee
From~(6) and (10a), (10b) it follows that operators $\widetilde x$, $\widetilde p$
satisfy the relations~(6) too, whereas the operators $(\widetilde S_{0a}, \widetilde S_{ab})$
defined by Eqs.~(10c) satisfy the commutation relations for the generators of the algebra
$O(1,n-1)$, as soon as the $S_{kl}$ satisfy the commutation relations for the generators
of the algebra $O(n)$.

Rewrite now the operators (3) in the form
\renewcommand{\theequation}{\arabic{equation}${}'$}
\setcounter{equation}{4}
\be
\ba{l}
P=(p_0, p_a, p_n),
\qquad
\ds J_{ab}=x_{[a} p_{b]} +S_{ab}, \qquad J_{0n}=x_{[a} p_{n]} +S_{an},
\vspace{2mm}\\
\ds J_{0a} = x_{[0} p_{a]} -\frac{S_{ab} p_b +S_{an} p_n}{p_0 +\sqrt{p_0^2-p_a^2 -p_n^2}},
 \qquad
J_{0n} = x_{[0} p_{n]} -\frac{S_{na} p_a}{\sqrt{p_0^2-p_a^2 -p_n^2}}.
\ea
\ee
Using the definitions (9) and (10) corresponding to the schematic substitution
``$i0\Leftrightarrow  n$'' when the operators with the symbol 2 ``$\sim $''
are getting from the operators without the symbol ``$\sim$'', we obtain from~(5):
\renewcommand{\theequation}{\arabic{equation}${}'$}
\setcounter{equation}{7}
\be
\ba{l}
P\equiv (\widetilde p_0, \widetilde p_a, \widetilde p_n)= (-ip_n, p_a, ip_0),
\vspace{2mm}\\
\ds \widetilde J_{ab}=\widetilde x_{[a} \widetilde p_{b]} +\widetilde S_{ab}, \qquad
i\widetilde J_{a0}=\widetilde x_{[a} \widetilde p_{0]} +i\widetilde S_{a0},
\vspace{2mm}\\
\ds -i\widetilde J_{na} = -i\widetilde x_{[n} \widetilde p_{a]} -
\frac{\widetilde S_{ab} \widetilde p_b -\widetilde S_{a0} \widetilde p_0}
{-i \widetilde p_n+\sqrt{-\widetilde p_n^2-\widetilde p_a^2 +\widetilde p_0^2}},
\vspace{2mm}\\
\ds \widetilde J_{n0} = \widetilde x_{[n} \widetilde p_{0]} -\frac{i\widetilde S_{0a} \widetilde p_a}
{-i\widetilde p_n+\sqrt{-\widetilde p_n^2-\widetilde p_a^2 +\widetilde p_0^2} }.
\ea
\ee
By virtue of Eqs. (10a), we have $\widetilde P^2=-P^2<0$, so that ($8'$) realizes a
representations of the class~III as soon as~(5) realizes a representation of the class~I.
Omitting in ($8'$) the symbol ``$\sim$'', we obtain (8).

Since all the representations of the class I are finite-dimensional, such a recipe allows to
obtain only finite-dimensional representations of the class~III (i.e., not all the representations
of this class). If, however, getting~(8) from~($8'$), the operators $(\widetilde S_{0a},
\widetilde S_{ab})$ will be substituted by operators $(S_{0a}, S_{ab})$
realizing an infinite-di\-men\-sio\-nal representation of the algebra $O(1, n-1)$,
we obtain the corresponding infinite-di\-mensional representation of the algebra
$P(1,n)$. Thus it is shown that the formula~(8) defines all the representations of the class~III
of the algebra $P(1,n)$.

The representations of the class~II ($P^2=0$, $P\not= 0$) requires a special treatment.
However, in the case when one of invariants of the algebra $P(1,n)$, namely, the invariant
\[
W\equiv \frac 12 P_\mu J_{\nu\alpha}^2 -P_\mu P_\nu J_{\mu \sigma} J_{\nu \sigma},
\]
vanishes, the representations of the class II are particular cases of representations of the
class~I, and formulaes for the generators $P_\mu$, $J_{\mu\nu}$ are obtained from~(5) by
the limit procedure $p^2\to 0$. The detailed discussion of all the representations of the class~II
is not given here. As to the class~IV $(P=0)$, in this case the group $P(1,n)$
reduces to the group $O(1,n)$, therefore the problem of classification and realization of
representations of the algebra $P(1,n)$ reduces to the problem of classification and
realization of representations of the algebra $O(1,n)$ already studied in ref.~[7].

Let us discuss now a role of the variable $x_0$. If we mean possibility to Interpret vectors
$\Psi$ constituting the representation space for the group $P(1,n)$,
as state vectors of the physical system (see below Section~3), we must interpret
$x_0$ as the time, i.e., as a parameter which is not an operator and which therefore is not
to be included in a complete set dynamical variables. It means that, for instance, in the
$x$-representation a vector-function $\Psi$ is a function of only $n$ dynamical variables:
$\Psi =\Psi(x_1,\ldots, x_n)$. If the condition~C of the Section~1 is not to be violated, the
number of independent dynamical variables in $p$-representation coincides with those in
$x$-representation, i.e., not all the dynamical variables  $p_0, p_1,\ldots, p_n$
are independent. For the representations in which the invariant $P^2$
is a fixed constant, the latter are connected by the relation
\renewcommand{\theequation}{\arabic{equation}}
\setcounter{equation}{10}
\be
p^2\equiv p_0^2-p_k^2=\varkappa^2>0, \qquad
p_0^2-p_k^2=-\eta^2<0
\ee
for the class I and III respectively. One can, for example, chose
\be
p_0 =\varepsilon \sqrt{p_k^2+\varkappa^2} \qquad \mbox{and}
\qquad p_0=\varepsilon \sqrt{p_k^2-\eta^2}, \qquad \varepsilon =\frac{p_0}{|p_0|}
\ee
for I and III. Then in $p$-representation  $\Psi=\widetilde \Psi(p_1, \ldots, p_n)$.
Of course, one can accept that $\Psi=\varphi(p_0,p_1,\ldots, p_n)$,
but under the condition~(11), so that
\be
\Psi=\varphi(p_0,p_1,\ldots, p_n) =\sqrt{2p_0}
\widetilde \Psi (p_1,\ldots, p_n) \delta(p^2-a^2), \qquad a=\varkappa^2,-\eta^2.
\ee

In the space of vector-functions $\Psi$ discussed the scalar product can be defined by
$P(1,n)$-invariant way:
\be
\ba{l}
\ds (\Psi, \widetilde \Psi) = \int d^{1+n} p\; \varphi^+(p_0,p_1, \ldots, p_n)
\varphi'(p_0, p_1, \ldots, p_n) =
\vspace{2mm}\\
\ds \qquad \qquad =\int d^n p\; \widetilde \Psi^+ (p_1, \ldots, p_n)
\widetilde \Psi' (p_1, \ldots, p_n).
\ea
\ee
The operators $P_\mu$, $J_{\mu\nu}$ defined in this space of vector-functions
$\Psi$,  have the form~(5) and~(8) where the sudstitution
\be
x_{[0}p_{k]}\to -\frac 12 (x_k p_0+p_0 x_k),
\ee
is made, $p_0$ is defined by (12),   $x_k$ and $p_k$ are defined by relations
\renewcommand{\theequation}{\arabic{equation}${}'$}
\setcounter{equation}{5}
\be
[x_k, p_l]=i\delta_{kl}, \qquad [x_k,x_l]=[p_k, p_l]=0,
\ee
while $S_{kl}$  and $(S_{0a}, S_{ab})$ are the same as in the formulae~(5) and~(8).

Thus, the ``quantum mechanical'' representation (of the Foldy--Shirokov~[3,~5] type)
of the generators  $P_\mu$, $J_{\mu\nu}$ of the algebra $P(1,n)$ is of the form:

For the class I
\renewcommand{\theequation}{\arabic{equation}}
\setcounter{equation}{15}
\be
\ba{l}
\ds P=(p_0, p_k), \qquad p_0\equiv \varepsilon \sqrt{p_k^2 +\varkappa^2},
\vspace{2mm}\\
\ds J_{kl}=x_{[k}p_{l]} +S_{kl}, \qquad
J_{0k} =-\frac 12 (x_k p_0 +p_0 x_k)-\frac {S_{kl}p_l}{p_0+\varkappa};
\ea
\ee

For the class III
\be
\ba{l}
\ds P=(p_0, p_k), \qquad p_0 \equiv \varepsilon \sqrt{p_k^2-\eta^2},
\vspace{2mm}\\
\ds J_{ab} =x_{[a}p_{b]} +S_{ab}, \qquad
J_{an} =x_{[a} p_{n]} -\frac{S_{ab} p_b -S_{a0} p_0}{p_n+\eta},
\vspace{2mm}\\
\ds J_{0n} =-\frac 12 (x_n p_0 +p_0 x_n) -\frac{S_{0a} p_a}{p_n +\eta},
\qquad
J_{0a} =-\frac 12 (x_a p_0 +p_0 x_a) +S_{0a}.
\ea
\ee

Since operators  $Q=x_k, p_\mu, S, P_\mu, J_{\mu\nu}$  of~(16) and~(17) are defined on
the apace of vectors $\Psi$ not depending on the time $x_0$, the representations~(16) and~(17)
are, in fact, the representations of the algebra $P(1,n)$ in the Heisenberg picture where for
the operators $Q$ as functions of the time $x_0$, the equation of motion
\be
i\p_0 Q =[Q, P_0]
\ee
is postulated.

In the Schr\"odinger picture vectors $\Psi$ depends explicitly on the time $x_0$
as on a parameter (but not as on a dynamical variable!) and for this dependence the equation
of the Schr\"odinger--Foldy type is postulated
\be
i\p_0 \Psi (x_0) =P_0 \Psi (x_0),
\ee
where  $P_0$ is defined by (12) and in $x$-representation $\Psi(x_0) =\Psi(x_0, x_1,\ldots, x_n)$,
in $p$-representation $\Psi(x_0)=-\varphi(x_0, p_0, p_1,\ldots, p_n)$
under the condition~(11) or $\Psi(x_0) =\Psi(x_0, p_1,\ldots, p_n)$ etc. These functions
are vector-functions, the manifold of which con\-stitutes the representation space for irreducible
representations of the group $P(1,n)$ in the Schr\"odinger picture. It is clear therefore that
their components are functions not only on $x_0, x_1,\ldots, x_n$ (or $x_0,p_1,\ldots,p_n$ etc.)
but also on ``spin'' variables $s_3,t_3, \ldots$ discussed above in connectian with
representations of homogeneous group $O(n)$ and $O(1,n-1)$. In accordance with the
domain of definition of ``spin'' variables $s_3,t_3,\ldots$ in different classes, the equation~(19)
is finite-component or infinite-component. In the class~I, where ``spin'' variables
$s_3,t_3,\ldots$ take only discrete and finite values, all the equations~(19) are
finite-component and their solutions $\Psi$ realises the unitary representations (i.e.,
vectors $\Psi$ are normalizable). In the class~III we have both finite-component
and infinite-component equations, but unitary representations can be realized only
on the solutions of the infinite-component equations.

One can suspect that owing to standing out of the time $x_0$ in the equation~(19), the last
is not invariant under the group $P(1,n)$ discussed. For the equation~(19) the conventional
demand of invariance under the given group is equivalent to the demand that the manifold
of its solutions is invariant under this group (i.e., that any of its solution under transformations
from $P(1,n)$ remains a solution of it but, generally speaking, another one). The mathematical
formulation of this requirement is to satisfy the condition
\be
[(i\p_0 -P_0),Q]\Psi=0,
\ee
where  $\Psi$ is any of solutions of Eq.~(19) and $Q$ is any generator of $P(1,n)$
or any linear combination of them, i.e., any element of the algebra $P(1,n)$.
Therefore the generators $Q=P_\mu, J_{\mu\nu}$ must have such a form that both the
relations~(4) and the condition~(20) must be satisfied. One can immediately verify that
such operators $P_\mu$, $J_{\mu\nu}$ are given by formulas~(5) and~(8) where, however,
the substitution
\renewcommand{\theequation}{\arabic{equation}${}'$}
\setcounter{equation}{14}
\be
x_{[0}p_{k]} \to x_0 p_k -\frac 12 (x_k p_0 +p_0 x_k)
\ee
is made and operators $x_k$, $p_\mu$ are defined by ($6'$) and (12).

Thus, the ``quantum mechanical'' representation of the generators $P_\mu$, $J_{\mu\nu}$
of the algebra $P(1,n)$ in the Schr\"odinger picture have the form:

For the class\footnote{Our formulae ($16'$) for generators $P_\mu$, $J_{\alpha\beta}$
in the case $P^2>0$ coinside with the corresponding formulae (B.25--28) in ref.~[5] if the
last are rewritten in the tensor form.} I
\renewcommand{\theequation}{\arabic{equation}${}'$}
\setcounter{equation}{15}
\be
\ba{l}
\ds P=(p_0, p_k), \qquad p_0= \varepsilon \sqrt{p_k^2 +\varkappa^2},
\vspace{2mm}\\
\ds J_{kl}=x_{[k}p_{l]} +S_{kl}, \qquad
J_{0k} =x_0 p_k-\frac 12 (x_k p_0 +p_0 x_k)-\frac {S_{kl}p_l}{p_0+\varkappa};
\ea
\ee

For the class III
\be
\ba{l}
\ds P=(p_0, p_k), \qquad p_0 = \varepsilon \sqrt{p_k^2-\eta^2},
\vspace{2mm}\\
\ds J_{ab} =x_{[a}p_{b]} +S_{ab}, \qquad
J_{an} =x_{[a} p_{n]} -\frac{S_{ab} p_b -S_{a0} p_0}{p_n+\eta},
\vspace{2mm}\\
\ds J_{0a} =x_0 p_a-\frac 12 (x_a p_0 +p_0 x_a) +S_{0a},
\vspace{2mm}\\
\ds  J_{0n} =x_0p_n-\frac 12 (x_n p_0 +p_0 x_n) -\frac{S_{0a} p_a}{p_n +\eta}.
\ea
\ee

It should be emphasized that in the Schr\"odinger picture the operators do not depend on
the time $x_0$, except of the operators $J_{0k}$. These last depend on the time $x_0$
only by due to the presence of the term $x_0 p_k$; it is just the presence of the term
$x_0p_k$ to satisfy the invariance condition~(20) of the equation~(19).

Note in the end of this section that last years the problem of using in physics some groups
like $P(m,n)$, $O(m,n)$ etc. as groups of generasized symmetry, was repeatedly arised
(see, for instance, ref.~[8] and refs. in ref.~[9]). The consequent physical analysis of a
quantumscheme based on either unificated group, is in fact possible only after a
mathematical analysis of representations of this group and equations connected with it, like
the analysis made here for the group $P(1,n)$. The method used here for studying the
representations of the group $P(1,n)$, is extend on the groups $P(m,n)$
without special difficulties. Thus the problem of classification of representations and
realization of an inhomogeneous group $P(m,n)$ is in fact reduced to the problem of
classification and realization of homogeneous groups of the type $O(m',n')$
already studied in ref.~[7].

\medskip

\centerline{\bf 3. Physical interpretation}

Last years many attempts of using different groups like $O(m,n)$, $P(m,n)$
as relativizing internal symmetry groups like $SU(n)$, were undertaken. The problem of
a relativistic generalization of an internal symmetry group is in fact connected with finding
a total symmetry group $G$ containing non-trivially the Poincar\'e group $P(1,3)$
(the group of ``external'' symmetry) and a group of ``internal'' symmetry like $SU(n)$.
As it is shown in refs.~[10], it is impossible non-trivially to unity the algebra $P(1,n)$
and the algebra of ``internal'' symmetries in the framework of a finite-dimensional algebra
Lie $G$, if the spectrum of the mass operator $M^2\equiv P_0^2-\vec P^{\,2}$
is discrete. In ref.~[11] a non-trivial example of the algebra $G \supset P(1,3)$
was constructed for the case when the spectrum of the mass operator is already stripe;
but the algebra $G$  was found to be infinite-dimensional in this case too. The consideration
of the infinite-dimensional algebras for the physical purposes is difficult both owing to
the absence of developed mathematical apparatus of such algebras and owing to
the necessity of solving a very difficult problem of physical interpretation of all the
commuting generators, the number of which is infinite. Do not speak about that the
question of writing down equations of motions, invariant under such an algebras, is
quite not clear. All this circumstances prompt that, to find a finite-dimensional algebra
$G \subset P(1,3)$ of a total symmetry group, we have to refuse from the demand of the
discreticity or even stripiticity of the mass spectrum. In this case one can propose many
groups as total symmetry groups (the groups of the type $P(m,n)$).
 However, in a $G=P(m,n), O(m,n)$, as like as in cases of other groups which are groups
of coordinate transformations in spaces of great dimensions, it still arises a difficult
problem of necessity to give a physical interpretation to the great number of
commuting operators.

Below we deal only with the inhomogeneous de Sitter group $P(1,4)$ which is a minimal
extention of the Poincar\'e group $P(1,3)$. Here we discuss a main topics of physical
interpretation of a quantum mechanical scheme based on this group. The group $P(1,4)$
is the most attractive because of in this case it is a success to give a clear physical
meaning to a complete set of commuting variables.

In $p$-representation a component of the wave function $\Psi$ ---  the a solution of the
equation~(19) with $n=4$ is a function of six dynamical variables of corresponding
complete set:
\[
\Psi(x_0, \vec p, p_4,s_3,t_3).
\]
As usually, this component is interpreted as the probability amplitude of finding
(by measuring at the given moment of the time $t=x_0$) the indicated values $\vec p$,
$p_4$, $s_3$, $t_3$ of the complete set $\vec P$, $P_4$, $S_3$, $T_3$.
The physical meaning of the operators $\vec P$ and $P_4$ is given in Section~1.
We discuss below the definition and physical meaning of the operators $S_3$, $T_3$
in the class~I.

Remind that in the case of $P(1,3)$ the operators $S_{kl}$ $k,l=1,2,3$  in ($16'$) which
constitute the spin vector $\vec S=(S_{23}, S_{31}, S_{12})$, are generators of the group
$O(3)$ (the little group of the group $P(1,3)$ in the class~I)
 and they are interpreted as an angular momenta of proper rotations.
More exactly they should be interpreted as an angular momenta which are connected with
intrinsic (internal) motion because when $\vec P=0$, the angular momenta $J_{kl}$
do not vanish but reduce to the spin angular monenta $S_{kl}$.

In the case of $P(1,4)$ there are six angular momentum operators, which describe
the internal motion of particle (i.e., the motion when $\vec p=p_4=0$):
$J_{kl/\vec p=p_4}=S_{kl}$, $k,l=1,\ldots,4$.
The operators $S_{kl}$ are generators of the group $O(4)$
(the little group of the group $P(1,4)$) in the class~I). They can be combined into two
3-dimensional vectors $\vec S$ and $\vec T$ defined by components
\renewcommand{\theequation}{\arabic{equation}}
\setcounter{equation}{20}
\be
S_a\equiv \frac 12 (S_{bc} +S_{4a}), \qquad T_a\equiv \frac 12 (S_{bc}- S_{4a}),
\ee
where $(a,b,c)=\mbox{cycl}(1,2,3)$. These components satisfy the relations
\be
\ba{l}
[S_a, S_b]=iS_c, \qquad [T_a, T_b]=iT_c,
\vspace{1mm}\\
\left[S_a, \vec S^{\,2}\right] =\left[T_a, \vec T^{\,2}\right]=[S_a, T_b]=0.
\ea
\ee

Remined that $\vec S^{\,2}$ and $\vec T^{\,2}$ are the invariants of the algebra
$O(4)$ being for irreducible representations $D(s,t)$ of this algebra
\be
\vec S^{\,2}=s(s+1)\hat 1, \qquad \vec T^{\,2}=t(t+1)\hat 1,
\ee
where $s,t=0,\frac 12, 1,\ldots$ and $\hat 1$  is the $(2s+1)(2s+1)$-dimensional unit matrix.
It was just the relations~(22) and~(23) to allow us~[1] to interpret 3-vectors
$\vec S$ and $\vec T$ as the spin and isospin operators.

It is clear from ($16'$) that in the representations of the class~I the generators of the algebra
$P(1,4)$  are constructed not only from the spin operators but also from the isospin
operators (and, of course, of $x_k$ and $p_k$). In this sense in quantum mechanic scheme
based on the group $P(1,4)$ the spin and isospin are presented on the same foot,
unlike from the case of conventional theory. Furthermore, unlike from to the latter, in our
case both the spin and isospin are entered dynamically. Indeed, in the conventional
approach the group $P(1,3)\otimes SU(2)_T$ is taken as the total symmetry group, so that
the generators of $SU(2)_T$ commute with the generators of $P(1,3)$  (even in the presence
of interactions). The group $P(1,4)$ which we taken as a total symmetry group, is not
isomorphic to the group $P(1,3)\otimes SU(2)_T$ furthermore, as in can be seen from~(21)
and ($16'$), $SU(2)_T \subset O(4)\subset P(1,4)$ as like as   $SU(2)_S \subset O(4)
\subset P(1,4)$, and the isospin operators (as like as the spin operators) do not commute with
$P(1,3) \subset P(1,4)$.

The manifold of solutions of the equation (19) realizes in the case discussed the irreducible
representation $D^\pm(s,t)$ of the algebra $P(1,4)$, where the sings ``$\pm$''
refer to the values $\varepsilon =\pm 1$ of the invariant --- the sign of energy, the numbers as
$s$ and $t$ determine the eigenvalues of the invariants
\be
\vec S^{\,2} =\frac{W}{4p^2} +\frac{V}{2\sqrt{p^2}}, \qquad
\vec T^{\,2} =\frac{W}{4p^2} -\frac{V}{2\sqrt{p^2}},
\ee
which are invariants both of  $P(1,4)$  and $O(4)$.

In quantum scheme based on $P(1,4)$,  possible states of an ``elementary particle''
(when  $\varepsilon =+1$) or ``antiparticle'' (when $\varepsilon =-1$)
 with given values of $s$, $t$ and $p^2=\varkappa^2$ are states which constitute the
representation space for an irreducible representation $D^\pm(s,t)$ of the group $P(1,4)$.
This is just the definition of the elementary particle in the $P(1,4)$-quantum scheme.
The simplest states of this particle are identified by eigenvalues of complete set of comuting
variables. It is clear that the representation  $D^\pm (s,t)$,
irreducible with respect to $P(1,4)$, is reducible with respect to $P(1,3)\subset P(1,4)$
therefore the ``elementary particle'' defined here, is not elementary in the conventional sense
(i.e., with respect to the group $P(1,3)$). Indeed, a solution $\Psi$ of Eq.~(19) with given
$s$ and $t$ contains componets identified not only by values of the 3-component
$s_3$ of spin but also by values of the 3-components  $t_3$, of isospin, so that the vector
$\Psi$ describes in fact the whole multiplet -- the set of states with different values of
$t_3$, $-t\leq t_3\leq t$ (and, of course, of $s_3$, $s\leq s_3\leq s$).
For example, the vector $\Psi^\pm$ with $\varepsilon \pm1$, $s=0$ and $t=\frac 12$
describes a meson isodublet like
\[
\Psi^+\equiv \left( \begin{array}{c}
\Psi^+_{(0,\frac 12)}
\vspace{1mm}\\
\Psi^+_{(0,-\frac 12)}
\end{array}\right)=
\left( \begin{array}{c}
K^+\\
K^0
\end{array} \right), \qquad
\Psi^-=\left( \begin{array}{c}
\widetilde K^0\\
K^-
\end{array}\right).
\]
The parameter $\varkappa$ (the threshold value of the free state energy or the ``bare''
rest mass) is the same for all the members of the given multiplet. Of course, the
introduction of a sutable interaction into the equation~(19) leads to the mass splitting
within a multiplet.

The approach proposed may be found successful for a consequent description of
unstable systems (resonances, particles or systems with non-fixed mass) already
in the framework of the quantum mechanics\footnote{The consequent consideration of
such problems demands, obviously, the quantum field approach, but a quantum
mechanical approach can be regarded as a half-fenomenology.}.
As it is known, the conventional quantum mechanical approach deals a with finding complex
eigenvalues of energy operators which must be Hermitian in a Hilbert space of wave
functions, i.e., in fact, one must go put of the framework of Hilbert space; the latter is
connected with breaking of such a fundamental principles as unitarity, hermiticity etc.~[12].

There are no similar difficulties in the quantum mechanical approach proposed. Indeed,
here the mass operator is an independent dynamical variable~(1), it is Her\-mi\-tian, defined in
the Hilbert space; therefore one can find its eigenvalues  $m^2$ and distributions
$\rho(m^2)$ in the same Hilbert space, as like as they find eigenvalues and distributions
for operators of energy, momentum and other dynamical variables. For example, if we have
a stationary wave function $\Psi=\left\{ \Psi(\vec x, x_4, s_3, t_3)\right\}$
of, generally speaking, an unstable multiplet (we meant: a solution of an equation of the
type~(19) with an interaction not depending on the time $x_0$) then
\be
\rho(m^2, s_3,t_3)=\int d^3 x\left| \int dx_4\; e^{-i\sqrt{m^2-\varkappa^2}x_4}
\Psi(\vec x, x_4, s_3, t_3)\right|^2.
\ee

If the distribution (25) with the given $s_3$, $t_3$ has one maximum, the experimentally
observed mass of the particle with given $s_3$, $t_3$ is defined either by the position
of the maximum or form
\be
\bar m^2 =\int d^3 x\; d x_4\; \Psi^* (\vec x, x_4, s_3, t_3) (p_4^2+\varkappa^2)
\Psi(\vec x,x_4, s_3, t_3)
\ee
and its mean lifetime  $\tau$ is defined from
\be
\bar m^2 \bar \tau^2 =1.
\ee
If the distribution (25) has more than one maximum, the position of them defines an
experimentally observed masses of unstable particles and the semi-widths of the
distribution~(25) in the regions of maximums define the lifetimes of corresponding
unstable particles. If, finally, $\rho(m^2,s_3,t_3)$ has a $\delta$-like singularity in a point
$m^2=m_0^2$, the $m_0$ is identified with the mass of a stable particle.

It is important to emphasize that in accord with our interpretation, the particles
experimentally observed are described not by the free equation~(19), but by an equation
of the type~(19) with a sutable interaction which may breakdown the $P(1,4)$-invariance,
but, of course conserves the $P(1,3)$-invariance\footnote{In this sense the consideration of
$P(1,4)$-symmetry here presented is only a base for its sutable violation --- analogously
to considerations and violations of $SU(n)$-symmetries.}.
As for solutions of the free equation~(19) they are some hypothetical (``bare'') states
which may not correspond, to any real particles. From view point of this interpretation
there are two types of interactions: interactions which cause a ``dressing'' of particles
and are inhierent even in asymptotical states, and usual interactions which cause a
scattering processes of real (``dressed'') particles. Therefore, in particular,
the 5-dimensional conservation law following from the free $P(1,4)$-invariant scheme;
may have not a real sense.

In the interpretation of the $P(1,4)$-scheme proposed we automatically have the
$SU(2)_T$-systematic  of particles. In contrast to the conventional systematics, our
one is a dynamical in the sense that for compaund model like those of
Fermi--Yang, Goldhaber--Gy\"orgyi--Kristy and others we can write down an equation
in which spin and isospin variables are mixed non-trivially.

Emphasize, that the interpretation of the $P(1,4)$-scheme proposed and, in par\-ti\-cu\-lar,
of the complete set of commuting variables mentioned above, was mainly based on
the definition~(1) of the varlable-mass operator as an independent dynamical variable.
This interpretation does not pretend, of course, to be the only one and complete.
In particular, the problem of giving the ``fifth coordinate'' $x_4$ the more immediate physical
sense than that one laying under its definition as a dynamical variable canonically
conjugated to the mass variable $p_4$, and the same problem refers to operators like
$J_{04}$, $J_{a4}$, $a=1,2,3$ we do not discusse here. The more detail discussion of these
problem is possible only in connection with considerations of solutions of equations like
Eq.~(19) with sutable interactions what is not a subject of this article.

Here we have considered the $P(1,n)$-invariant equations of the Schr\"odinger--Foldy
type in an arbitrary dimensional Minkowski space, in which the differential operators
$\p_k\equiv \p/\p x_k$ of the ``space'' variables are presented under square root.
This equations describing the positive and negative states separatelly, are suitable
for quasirelativistic quantum mechanical considerations (e.g., for calculations of
spin-isospin effects in $P(1,4)$-invariant equations with interactions included).
Theoretic-field considerations are usually based on equations of first order on $\p_\mu$.
The general form of $P(1,n)$-invariant linear on $\p_\mu$ equation is
\be
(B_\mu\p_\mu \pm \varkappa)\Phi^\pm =0, \qquad \mu=1,2,\ldots,n,n+1,
\ee
where the operators $B_\mu$ are defined by the relations
\be
[B_\mu, J_{\rho\sigma}]=\delta_{\mu\rho}B_\sigma- \delta_{\mu\sigma} B_\rho,
\qquad (\mu,\rho, \sigma=1,\ldots, n+1).
\ee
For the representation of the class I the operators $B_\mu$ are finite-dimensional;
for those of the class~III the operators $B_\mu$ can be both finite- and infinite-dimensional.
Concrete forms of operators $B_\mu$ can be founed by the method proposed in ref.~[13].

In this paper we have not considered the problem of invariance of the equation~(28) as to
discrete transformations, that is relatively to
\be
x'_k=-x_k, \qquad x'_0=-x_0.
\ee
This problem has been investigated by one of us~[14]. As it is shown in~[14] the equa\-tion~(28)
for $n=2m$, $m=1,2,3,\ldots$, is neither invariant as to transforma\-tions~(30) nor
\be
x'_0=-x_0, \qquad x'_k=x_k, \qquad k=1,2,\ldots, 2m.
\ee

Thus, in the field theory constructed on the basis of the groups $P(1,2)$, $P(1,4)$,
$P(1,6)$ and so on the theorem $CPT$ may be broken down. It should be emphasized,
however, that the direct of the manifold of solutions $\{\Phi^+\}$ and $\{\Phi^-\}$
$T$-, $CPT$-invariant.

\medskip

\begin{enumerate}
\footnotesize

\item Fushchych W.I., Krivsky I.Yu., {\it  Nucl. Phys. B}, 1968, {\bf 7}, 79;\ \ {\tt quant-ph/0206057} \\
Fushchych W.I., Krivsky I.Yu.,  Preprint ITF-68-72, Kyiv, 1968.

\item  Wigner E.P., {\it  Ann. Math.}, 1939, {\bf 40}, 149.

\item  Shirokov Yu.M., {\it JETP (Sov. Phys.)} 1957, {\bf 33}, 1196.

\item Joos H., {\it Fortschr. Phys.}, 1962, {\bf 10}, 65.

\item Foldy L., {\it  Phys. Rev.}, 1956, {\bf 102}, 568.

\item Gelfand I.M., Zeitlin M., {\it DAN USSR}, 1950, {\bf  71}, 1017.

\item Gelfand I.M., Grayev M.I., {\it Trudy Moskovskogo Matematicheskogo obshchestva},
1959, {\bf  8}; {\it Izv. AN USSR, Ser. Math.}, 1965, {\bf  29},  5.

\item Hederfledt G.C., Henning J., {\it Fortschr. Phys.}, 1968, {\bf 16}, 491.

\item Sokolik G.A., Group methods in theory of elementary particles,
Atomizdat, Moscow, 1965.

\item O'Raifeartaigh L., {\it Phys. Rev. Lett.}, 1965, {\bf 14}, 575; \\
Jost R., {\it Helv. Phys. Acta}, 1966, {\bf 39}, 369.

\item Fushchych W.I., {\it Ukrain. Phys. J.},  1968, {\bf 13},
256.\ \ {\tt quant-ph/0206056}

\item Mathews P.T., Salam A., {\it Phys. Rev.}, 1958, {\bf 112}, 283.

\item  Fushchych W.I., {\it Ukrain. Phys. J.}, 1966, {\bf 8}, 907.

\item Fushchych W.I., {\it  JETP Letters}, 1969.

\end{enumerate}

\end{document}